\documentclass[12pt,letterpaper]{article}

\usepackage[left = 1 in, right = 1 in, top = 1 in, bottom = 1 in]{geometry}

\usepackage{amsmath}
\usepackage{amssymb} 
\usepackage{graphicx}
\usepackage{setspace}
\usepackage{float}        
\usepackage{siunitx}
\usepackage{xspace}
\usepackage{xcolor}
\usepackage{listings} 

\newcommand{\refcite}[1]{Ref.~\cite{#1}}
\newcommand{\sioo}[0]{SiO\textsubscript{2}\xspace}

\newcommand{\Ohm}[0]{$\Omega$\xspace}

\newcommand{\uA}[0]{\textmu A\xspace}

\begin{document}

\doublespacing
\pagenumbering{arabic} 







\title{The thermally-coupled imager: A scalable readout architecture for superconducting nanowire single photon detectors}

\author{A. N. McCaughan$^1$, Y. Zhai$^1$, B. Korzh$^2$, J. P. Allmaras$^2$, \\ B. G. Oripov$^1$,  M. D. Shaw$^2$, \& S. W. Nam$^1$}

\date{
    \small
    $^1$National Institute of Standards and Technology, Boulder, CO 80305\\%
    $^2$Jet Propulsion Laboratory, California Institute of Technology, 4800 Oak Grove Dr., Pasadena, CA, USA
}
\maketitle


\section*{Abstract}

Although superconducting nanowire single-photon detectors (SNSPDs) are a promising technology for quantum optics, metrology, and astronomy, they currently lack a readout architecture that is scalable to the megapixel regime and beyond.  In this work, we have designed and demonstrated such an architecture for SNSPDs, called the thermally-coupled imager (TCI).  The TCI uses a combination of time-of-flight delay lines and thermal coupling to create a scalable architecture that can scale to large array sizes, allows neighboring detectors to operate independently, and requires only four microwave readout lines to operate no matter the size of the array.  We give an overview of how the architecture functions, and demonstrate a  proof-of-concept $32\times32$ imaging array. The array was able to image a free-space focused spot at 373 nm, count at 9.6 Mcps, and resolve photon location with greater than 99.83\% distinguishability.

\section{Introduction}

Currently, superconducting-nanowire single-photon detectors (SNSPD) are one of the most promising single-photon detection technologies for quantum applications and metrology. However, many of the present and future applications for these single-photon detectors require large arrays which presently do not exist~\cite{McCaughan2018}. For example, imaging multi-photon spatial correlations in quantum optics would greatly benefit from having large, high-efficiency arrays~\cite{Edgar2012}. Of particular interest is also the application of SNSPDs to astronomy, particularly in the ultraviolet (UV); presently, there are only a few large-scale detector technologies available in the wavelength range of 90-400 nm. Future high-precision UV astronomy and astrophysics observations require 100-megapixel-scale arrays of detectors which are single-photon sensitive, solar-blind, and have high detection efficiencies. 

To date, there has not been a multiplexing architecture capable of scaling SNSPD arrays to megapixel sizes and larger. Previous approaches have pursued a variety of approaches, including row-column readouts~\cite{Wollman2019}, frequency-multiplexing, SFQ-based readout, and time-of-flight imaging~\cite{zhao2017single}. However, the best array architectures to date have only managed pixel counts on the order of $\sim$1,000 pixels, and their architectures are unlikely to extend beyond the 10,000 pixel count due to signal-to-noise limitations~\cite{Allman2015}. These limitations are due to the difficulty of multiplexing the low-amplitude, broadband output of the individual detectors~\cite{McCaughan2018}. These types of signals are not easily multiplexed onto microwave readout lines and, since the detectors are fundamentally cryogenic, there are practical cooling-power limits to how many readout lines can be used to carry signals from the array to room-temperature readout electronics. An ideal multiplexing scheme should be able to (1) scale to large array sizes, (2) allow neighboring detectors to operate independently (minimize crosstalk), and (3) ideally use only a few readout lines.

\begin{figure}[ht] 
    \centering
    \includegraphics[width=6.5in]{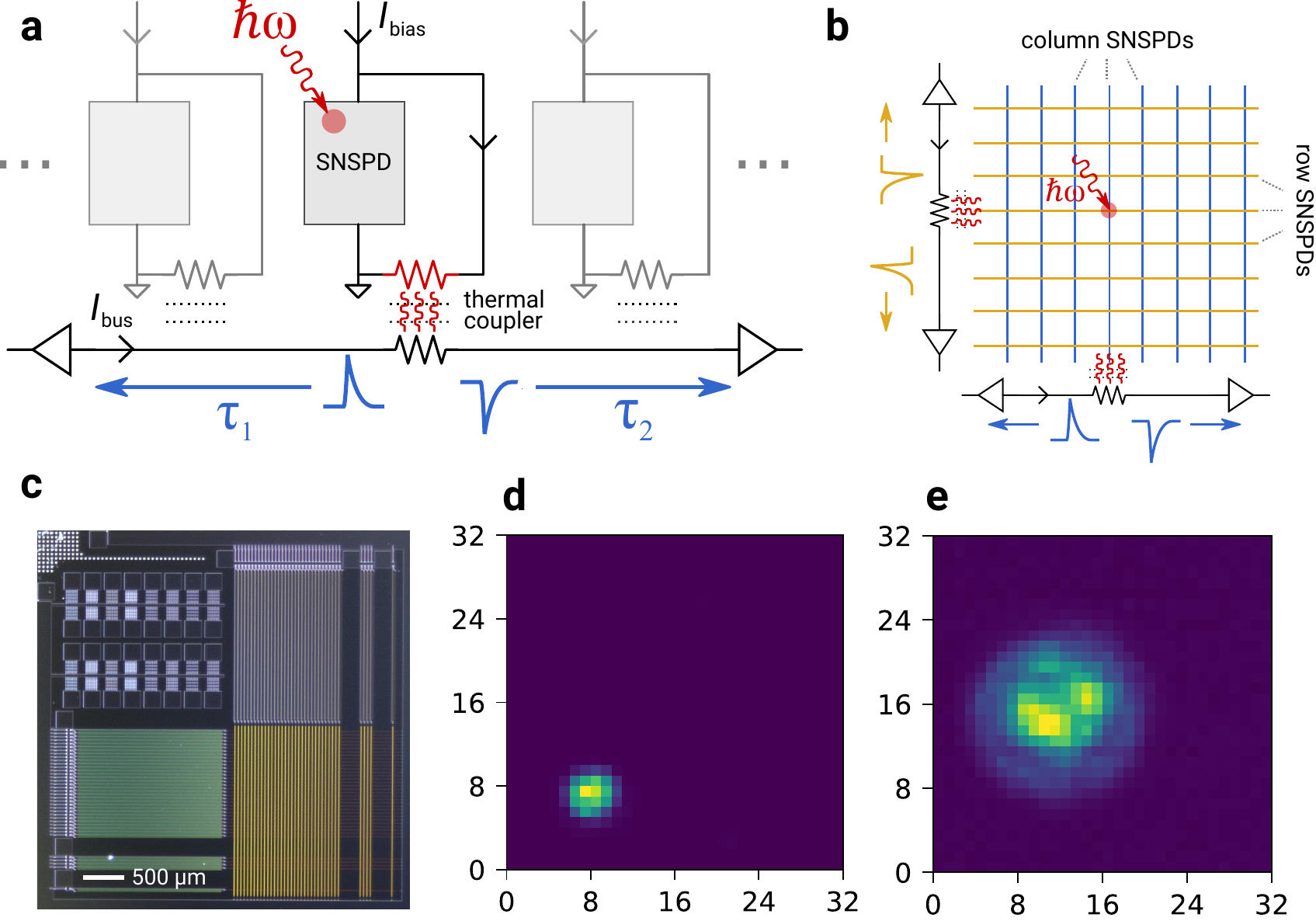}
    \caption{Overview of the kilopixel thermally-coupled imager(TCI) demonstrated here.  (a) Diagram of the TCI array operation in one dimension.(b) Constructing a two-dimensional imager from two sets of one-dimensional arrays overlapped at right angles. (c) Microscope image of the fabricated $32\times32$ imager.  (d, e) Imaging a 373 nm spot with varying focus.}
    \label{fig1}
\end{figure}

Here we present the "thermally-coupled imager" (TCI) multiplexing architecture that achieves all of these goals. The unique aspect of this architecture compared to previous work is that each detector is independently thermally-coupled to a readout bus.  By using thermal-coupling instead of electrical coupling, scalability issues due to electrical crosstalk and signal absorption are drastically reduced. Instead, the tradeoff made is one of time; the readout bus uses an electrical time-of-flight delay mechanism to derive the position of the photon, and larger array sizes require longer time-of-flight delays. As a result, scaling the array to larger pixel counts does not degrade the distinguishability of individual detectors or reduce signal-to-noise of any individual detector -- it only reduces the maximum readout rate (as is the case for most imaging technologies such as CCDs and CMOS sensors). Additionally, reading out an entire array requires only 4 microwave lines, regardless of the size of the array.

\section{Background}

The most successful SNSPD-multiplexing architecture to date used passive electrical coupling~\cite{Wollman2019}, and demonstrated 1024-pixel array sizes. This style of multiplexing used a resistive network to distribute output signals row-column readout lines. In this architecture and related ones, multiplexing is done with passive electrical components, meaning the coupling from the detector to the readout is necessarily symmetric to the coupling from the readout to the detector. As a result of this symmetric coupling, when multiple detectors share a common readout line, an output signal generated from any one detector is partially absorbed by its neighbors.  This absorption limits the ultimate scalability of this type of architecture; as more neighboring detectors are added, more of the output signal is inadvertently lost until it eventually falls below the noise level of the readout and the output signal becomes unrecoverable.

Another type of passive architecture avoids the symmetric-coupling problem by assigning each detector a unique frequency~\cite{Doerner2017, Sinclair2019}. These RF-multiplexing schemes can make excellent use of the bandwidth avaliable on a microwave readout line, but use large amounts of homodyne readout circuity at room temperature to sort through the multiple resonances, and require each resonance to be fine-tuned so that the homodyne circuitry can operate correctly.  There are also active readout architectures that are potentially scalable, including nTron-based approaches~\cite{Zheng2019} and SFQ-based approaches~\cite{Hofherr2012,Miyajima2018,Yamashita2012}, but to date the largest arrays shown in these systems have been significantly smaller than passive approaches~\cite{Gaggero2019}.

The direct antecedents to the TCI architecture are the SNSPI~\cite{zhao2017single} and the thermal row-column array~\cite{Allmaras2020}. The SNSPI used a single nanowire to both detect photons and resolve their position via time-of-flight. This architecture was able to achieve an effective pixel count of 590, but its single-nanowire nature has practical drawbacks that make it unlikely to be scalable to megapixel regimes. The thermal row-column array utilizes thermal-coupling to allow a single photon to trigger two different detectors (a row and column detector), but does not multiplex the signals in any other way, meaning 2,000 microwave readout lines would be required to achieve a megapixel array. The TCI architecture utilizes the best advantages from both of these schemes, and avoids the many of the pitfalls both as well.

\section{Architecture description}

The TCI architecture is composed of 3 primary components: (1) individual detectors, (2) thermal couplers, and (3) a time-of-flight readout bus. The detectors take the form of standard current-biased SNSPDs. Each thermal coupler consists of a resistive heater-element that is thermally coupled to (but electrically isolated from) the readout bus by a thin (25 nm) electrically-insulating \sioo spacer. The heater-element of the thermal coupler can either be a resistive material, or a weakly-superconducting material with a low switching current that becomes resistive when current is forced through it.  For the work presented here, it is the latter. Similar to the SNSPDs, the time-of-flight readout bus is also a wire made from a superconducting thin-film that is current-biased.

While the primary results of this work are for a two-dimensional array, we first explain the operation in one dimension for clarity. As shown in Fig. 1a, the detectors are arranged in a line along the length of the readout bus. At rest, all the SNSPDs and the readout bus are fully superconducting, current-biased, and produce no voltage transients, and the entire device is at a uniform cryogenic temperature. When a photon is detected on one of the SNSPDs, current is diverted from the detector into the input of the thermal coupler. 

Phonons generated from the Joule heating in the heater-element are then coupled through the dielectric layer of the thermal coupler and into the nanowire readout bus, similar to the process described in~\refcite{McCaughan2019b}. These phonons are then absorbed by the superconducting readout bus, locally weakening the superconductivity and generating a resistive hotspot in the bus. Due to the readout bus bias-current $I_{bus}$, the resistive hotspot generated in the readout bus then produces two counter-propagating readout pulses. As shown in Fig. 1a, a positive voltage pulse travels left, while a negative voltage pulse travels right. Critically, these readout pulses do not interact with any of the neighboring detectors while they propagate; besides a negligible ($<$ 1 fF) capacitance, the only available interaction between the detector and the readout bus is thermal.  As a result, the neighboring detectors on the readout do not absorb any meaningful quantity of the generated output signal.

Once the pulses are created on the readout at time $t_p$, the architecture then uses a time-of-flight method similar to \refcite{zhao2017single} to determine the hotspot location. Due the extremely high kinetic inductance of the readout superconducting nanowire, these voltage pulses travel the length $L$ of the readout bus at a velocity $v$, a small fraction of the speed of light, until they are detected by the amplifiers at either end of the readout bus. From these four quantities, we can determine the position of the pulse origin $x_p=((\tau_2-\tau_1 )v+L)/2$, and the time of pulse creation from $t_p=((\tau_2+\tau_1 )-L/v)/2$. Provided there is sufficient time-spacing on the bus between adjacent detectors, we can then determine which detector received absorbed the photon.  We note, however, that only one detection is possible per readout time-of-flight period without causing disambiguation issues.

To create a 2D imager, then, we overlap two of the 1D arrays at right angles such that one set of detectors corresponds to rows, and the other corresponds to columns as shown in Fig. 1b. Physically, the row detectors and column detectors are stacked vertically, separated by 25 nm of \sioo, in a configuration similar to the one presented in \refcite{Allmaras2020}. When a photon arrives, it is absorbed into one detector layer (e.g. a row detector) and creates a hotspot. The hotspot then creates phonons that that travel a short distance through the dielectric spacer and impinge upon the other detector layer (e.g. a column detector), triggering it. Thus, two detectors are triggered (one row, one column), and their respective readout busses are used to determine the row and column in which the photon was detected. Using this row-column detector scheme minimizes the number of total detectors required, and additionally reduces the overall length of the readout bus.

\section{Results}

We implemented the TCI architecture using WSi-based SNSPDs and characterized a 32x32 array, resolving 1024 pixels. The active area contains only the two layers of detectors--the ancillary circuitry needed to perform the multiplexed readout is located along the edges of the active area.  The SNSPDs were 1-um-wide WSi wires, and the thermal coupler inputs were 350-nm-wide constrictions made from WSi.  The readout bus was also constructed from a 1.5-um-wide WSi wire.  The array was measured at 0.98~K in a cryostat, and light was free-space coupled to the array from room temperature through a vacuum window. Shown in Fig. 1d is the imaging of a 373~nm laser spot. As seen in Fig. 2, we observed saturation in the detection efficiency at 373 nm, indicating near-unity quantum efficiency, and were able to count at a rate of 9.6 Mcps.

\begin{figure}[ht] 
    \centering
    \includegraphics[width=3.5in]{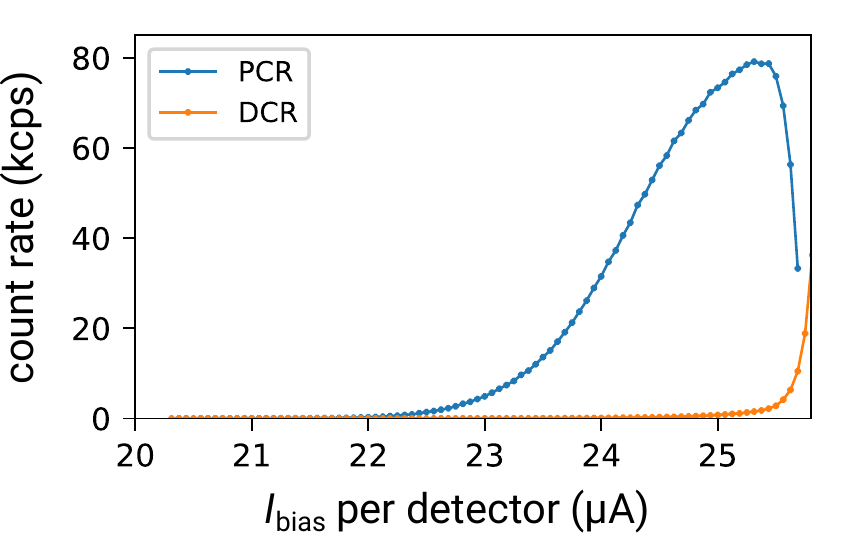}
    \caption{Photon count rate at 373~nm (blue) and dark count rate (orange) in the $32\times32$ array. }
    \label{fig2}
\end{figure}

During operation, the full-width half-max of the readout jitter was 37.4 ps (std. dev. $\sigma$ of 15.9 ps), well below the 379 ps time-of-flight spacing between each detector ($23.8 \sigma$ separation). To measure the uniformity of the readout process, we also performed a flood illumination of the array, shown in Fig. 3. Measuring 320,000 total counts, we found a lower bound of 99.83\% distinguishability between adjacent rows/columns.  Given the 23.8-$\sigma$ separation between adjacent detectors in the readout, we suspect this is likely even closer to unity, but external noise may have unnecessarily corrupted a handful of counts. In this initial demonstration, the fill factor was very low, approximately 0.5\%.  In future work, however, this value could be increased to near-unity, as the only elements within the imaging area are the photosensitive nanowires.

\begin{figure}[ht] 
    \centering
    \includegraphics[width=6.5in]{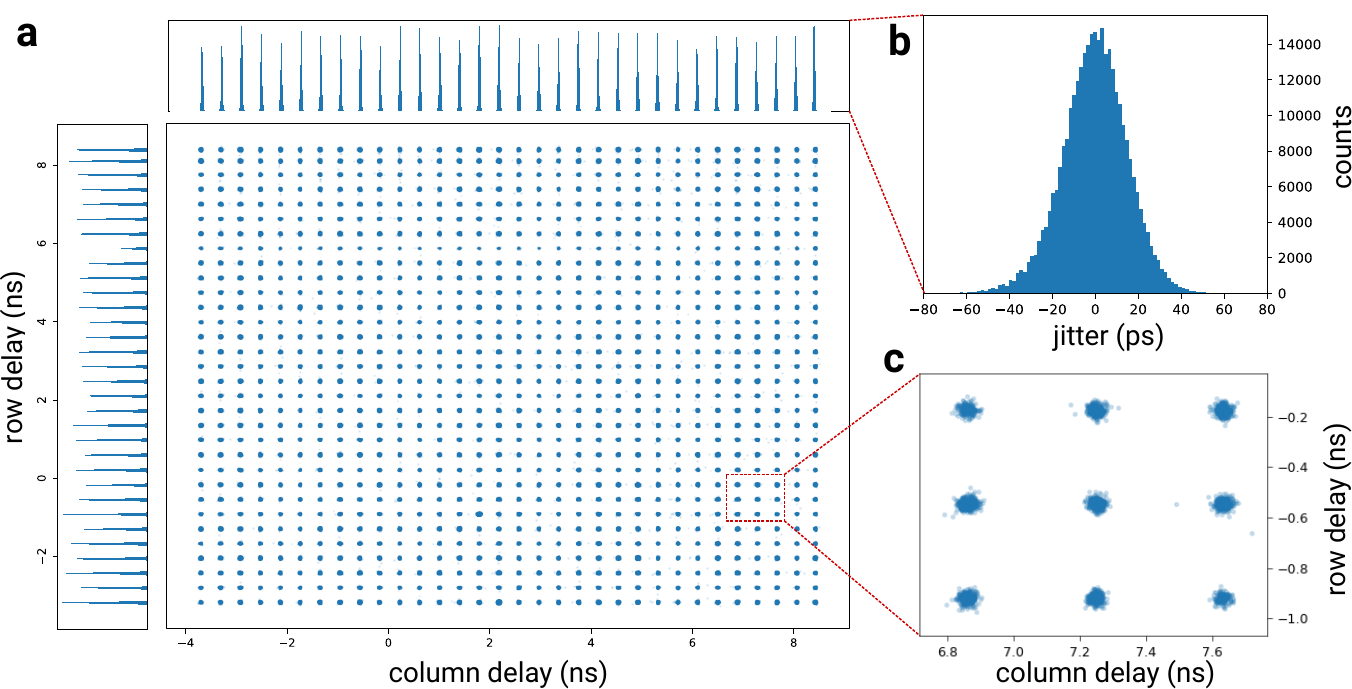}
    \caption{Measuring the distinguishability of the readout mechanism by uniform illumination of the $32\times32$ array.  (a) Scatter plot of the measured differential time-delay ($\tau_1$-$\tau_2$) for the row and column readout busses for 320,000 counts. Shown on the sides are histograms for the row and column busses. (b) Summation of all column histogram bins, modulo 379 ps, into a single histogram showing a standard deviation of 15.9 ps. (c) Zoom-in of the differential time-delay scatter plot.}
    \label{fig3}
\end{figure}

Current was distributed to individual detectors via an on-chip resistive distribution network, using 12.2 \Ohm/sq gold-palladium for the resistors and 90-nm-thick niobium for the wiring. As shown in Fig. 4, each detector has a support circuit with a series resistor $R_{series}$ = 130.1 \Ohm, shunt resistor $R_{sh}$ = 24.4 \Ohm, and thermal coupler resistor $R_{tc}$ = 24.4 \Ohm. For the results shown in Fig. 1d, a total of 750 \uA current was injected to each set of 32 detectors, meaning each 1-um-wide detector carried 23.4 \uA of bias current.  The detectors were not biased fully near saturation due to latching issues with the readout, likely caused by impedance mismatches in the tapered readout. 

\begin{figure}[ht] 
    \centering
    \includegraphics[width=3.5in]{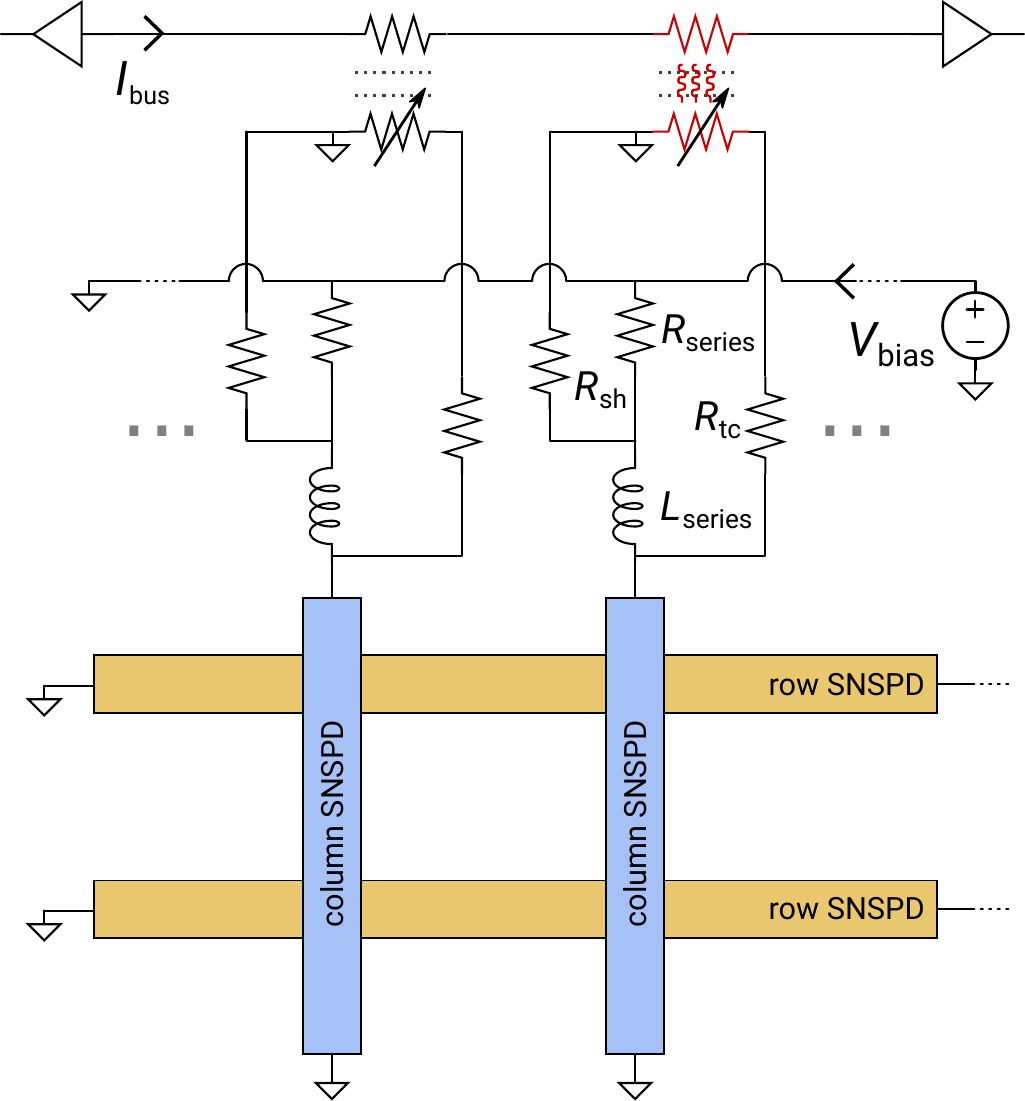}
    \caption{Circuit diagram showing the configuration and approximate layout of the ancillary support-circuitry used in the TCI}
    \label{fig4}
\end{figure}

To guarantee operation, we performed extensive tests on the readout circuitry during the design phase. Measurement of thermal-coupler test structures indicated that only 240 eV of electrical energy was required on the thermal coupler to trigger the output from the readout bus.  In the array, this thermal-coupling energy was stored in large kinetic-inductance-based nanowire series inductors ($L_{series}$ in Fig. 4). When a photon first created a resistive hotspot in a detector nanowire, $L_{series}$ acted as a stiff current-source, guaranteeing that most of the current from the detector was forced into the thermal coupler and did not parasitically leak into neighboring detectors through the resistive bias network. Testing showed that a $L_{series}$ value of 1~uH was sufficient to guarantee that, for every photon detection, enough current was diverted into the thermal coupler to generate a corresponding output on the readout.

The pulses that came out of the readout bus did not show any noticeable signal degradation, even in test structures where we scaled the number of detectors to 516. Due to the large kinetic inductance of the nanowire, the propagation velocity along the readout bus was 0.0069*c (2.07 um/ps). Hecken tapers were placed on the ends of the readout bus to prevent internal reflection of the fast rising edge from the readout pulses~\cite{zhao2017single}.

\section{Conclusion}
In summary, we have designed and demonstrated an SNSPD architecture that is scalable to the megapixel and beyond.  The imager design was targeted for UV-astronomy, and showed single-photon sensitivity and unity internal quantum efficiency across 1024 pixels at 373 nm. Crucially, the architecture itself is wavelength-agnostic, and should be applicable to any wavelength SNSPDs are capable of detecting (UV~\cite{Wollman2019} to mid-IR~\cite{verma2021single}).

However, several areas for improvement remain. The most pressing next step is to build bigger arrays and demonstrate a true single-photon megapixel imager.  Also, in the present design, the array only has a 0.5\% fill factor, which is far too low for many astronomical and quantum-optical applications. Fortunately, because all the readout circuitry can be located outside of the imaging area, the architecture has a direct path to achieving unity fill-factors. Other steps forward include further analysis of count-rate limitations to scalability, since the size of the array affects the dead-time between consecutive photon detections. We note the addition of multiple readout busses allow for a favorable quadratic scaling effect to the count-rate -- for instance, by doubling the number of readout busses, the number of detectors (and thus length of each bus) is reduced by half and the number of photons that can be read out per dead-time is doubled.

\section{Acknowledgements}

The U.S. Government is authorized to reproduce and distribute reprints for governmental purposes notwithstanding any copyright annotation thereon. Part of this research was performed at the Jet Propulsion Laboratory, California Institute of Technology, under contract with NASA. A.N.M. was supported in part by NASA APRA grant NNH17ZDA001N. J.P.A. acknowledges support from a NASA Space Technology Research (NSTRF) fellowship.





\newpage

\providecommand{\noopsort}[1]{}\providecommand{\singleletter}[1]{#1}%

\end{document}